\UseRawInputEncoding
\documentclass[lettersize,journal]{IEEEtran}
\usepackage[utf8]{inputenc}
\usepackage[T1]{fontenc}
\usepackage{amsmath,amsfonts}
\usepackage{algorithmic}
\usepackage{algorithm}
\usepackage{array}
\usepackage[caption=false,font=normalsize,labelfont=sf,textfont=sf]{subfig}
\usepackage{textcomp}
\usepackage{stfloats}
\usepackage{url}
\usepackage{verbatim}
\usepackage{graphicx}
\usepackage{cite}

\begin{document}

\title{Integrating Identity-Based Identification against Adaptive Adversaries in Federated Learning}

\author{Jakub Kacper Szel\k{a}g, Ji-Jian Chin, Lauren Ansell, Sook-Chin Yip*\\
\thanks{Jakub Kacper Szel\k{a}g, Ji-Jian Chin and Lauren Ansell are with School of Engineering, Computing and Mathematics, University of Plymouth, Drake Circus, Plymouth, PL4
8AA, United Kingdom (jakub.szelag@students.plymouth.ac.uk; ji-jian.chin@plymouth.ac.uk; lauren.adams@plymouth.ac.uk)}
\thanks{Sook-Chin Yip is with Faculty of Engineering, Multimedia University, Persiaran Multimedia, Cyberjaya, 63100, Malaysia (scyip@mmu.edu.my)}
\thanks{The source code for our solution and experiments may be found on https://github.com/SzelagJK/IBFL.git}
\thanks{A supplementary video for this work may be found on https://youtu.be/Q-JDm0NU6cQ}
\thanks{This work has been submitted to the IEEE for possible publication. Copyright may be transferred without notice, after which this version may no longer be accessible.}
\thanks{*Corresponding author: scyip@mmu.edu.my}}

\markboth{}%
{}

\IEEEpubid{}

\maketitle

\begin{abstract}
Federated Learning (FL) has recently emerged as a promising paradigm for privacy-preserving, distributed machine learning. However, FL systems face significant security threats, particularly from adaptive adversaries capable of modifying their attack strategies to evade detection. One such threat is the presence of Reconnecting Malicious Clients (RMCs), which exploit FL’s open connectivity by reconnecting to the system with modified attack strategies. To address this vulnerability, we propose integration of Identity-Based Identification (IBI) as a security measure within FL environments. By leveraging IBI, we enable FL systems to authenticate clients based on cryptographic identity schemes, effectively preventing previously disconnected malicious clients from re-entering the system. Our approach is implemented using the TNC-IBI (Tan-Ng-Chin) scheme over elliptic curves to ensure computational efficiency, particularly in resource-constrained environments like Internet of Things (IoT). Experimental results demonstrate that integrating IBI with secure aggregation algorithms, such as Krum and Trimmed Mean, significantly improves FL robustness by mitigating the impact of RMCs. We further discuss the broader implications of IBI in FL security, highlighting research directions for adaptive adversary detection, reputation-based mechanisms, and the applicability of identity-based cryptographic frameworks in decentralized FL architectures. Our findings advocate for a holistic approach to FL security, emphasizing the necessity of proactive defence strategies against evolving adaptive adversarial threats.
\end{abstract}


\section{Introduction}
In recent years, Federated Learning (FL) has emerged as one of the leading paradigms for distributed machine learning \cite{mcmahan_communication-efficient_2017}. One of its main goals is to significantly improve privacy through data decentralization, where each client is in possession of its own dataset. In short, FL operates around the idea of training smaller local models on individual clients (without sharing their local datasets) with purpose of later aggregation of those models that will produce a global model. Besides theoretical improvements to privacy-preservation of machine learning processes, FL equally aims to significantly reduce the reliance of machine learning on High-Performance Computation (HPC) \cite{annunziata_dynamics_2024}. This is especially relevant to fields such as Internet of Things (IoT) \cite{zhang_federated_2022}, vehicular networks\cite{elbir_federated_2022}, and healthcare \cite{chaddad_federated_2024, nguyen_federated_2022} as limited computational capabilities of individual devices presents itself to be one of the main limitations for machine learning applications. As such, we may observe a regular stream of ideas form the scientific community that attempt to utilise FL, allowing them to generate models with accuracy close to that of centralised models. \par

Acknowledging the plethora of applications as well as their individual architectures, FL further evolved to accommodate for those scenarios. This also applies to the data distribution on individual clients as well as centralisation or decentralization of the aggregation process itself. Effectively proving the potential FL holds for such and similar adaptations, as recent works \cite{castiglia_flexible_2024, zeng_fedlab_2023, xie_federatedscope_2022} show the flexibility that FL provides with its novel approach.\par

Nevertheless, despite the initial hopes of improvements to privacy-preservation, it is known to researchers that FL faces a variety of threats to not only privacy-preservation but also model-robustness \cite{xie_survey_2024, almutairi_federated_2023}. To be specific, FL has opened a gate to the prevalence of a variety of model-poisoning attacks aiming to degrade global models performance \cite{zhou_deep_2021}, as well as data-reconstruction and model-inversion attacks targeting the privacy of clients datasets \cite{liang_egia_2023, zhao_loki_2024}. On top of that, the fact that the local models must be transferred from the clients to either other clients or the aggregator means that attacks against privacy and integrity of FL models may be both external and internal. Consequently, this has prompted the research community to further advance the field of FL security such that the aforementioned threats would have been prevented. To list a few, some of the leading defensive mechanisms adopted by researchers in FL security are: (1) Secure Aggregation \cite{liu_privacy-preserving_2022}, which in most cases attempts to use statistical approaches in order to filter the poisoned local models out of the aggregation such that the global model is left unaffected, (2) Homomorphic Encryption \cite{park_privacy-preserving_2022},  allowing for mathematical operations to be performed over ciphertext such that only the client will have access to the plaintext, (3) Differential Privacy \cite{ouadrhiri_differential_2022}, introducing noise to the datasets in a strategic manner such that it is difficult to tell for an adversary if records are present within clients dataset, and (4) Dynamic Making \cite{zhang_uncovering_2024}, aiming to obfuscate local model updates in a way that an adversary would not be able to perform reconstruction attacks even if they have access to observe such updates.\par

In spite of that, one of more recently established methods of attack has introduced involvement of adaptive adversaries (AAs) which are capable of modifying their behaviour during the attack and subsequently adapt to the defensive measures acting against their attacks \cite{kraus_automatic_2024, bagdasaryan_how_2020}. This is to maximise the severity of attacks in relation to the adversaries aim while minimising the chances of defensive measures stopping adversaries from doing so. An example of that serves Reconnecting Malicious Clients (RMCs) \cite{szelag_adaptive_2025} which have been recently noticed to have a significant potential as a form of adaptive adversarial attack. Assuming FL environment allows for clients to reconnect without the system keeping track in regard to who is reconnecting with effective ways of blocking connections, RMCs may repetitively try to attack systems integrity and privacy by simply reconnecting back with another attack strategy. This paper concentrates on this security flaw within FL systems. \par

We highlight that the potential solution we aim to explore against this threat may lie with Identity-Based Identification (IBI) schemes, as the underlying cryptographic primitives for IBI are directly reliant on the identities of involved parties. Intuitively, this presents itself as one of the natural developments for security within FL addressing this issue. The general idea is to use IBI along with clients identities for the system to authenticate individual clients as well as keep track of the disconnected parties to ensure that any malicious clients will not be allowed back into the system once forcibly disconnected. We notice that this approach may prove to be effective especially in cloud environments where clients identities are more difficult for the adversary to change or replace depending on the security configurations \cite{lora_applying_2024}. As such, this applies to all similar systems where this assumption holds true. Further, we acknowledge that forcible disconnections from the system necessitates use of detection systems, henceforth our presented approach will concentrate on implementations over secure aggregation algorithms.\par

However, it needs to be noted that the domain of FL security is still at its early stages. Solutions we have so far provided as a community are often put in isolated environments targeting individual threats \cite{zhang_security_2022, wen_survey_2023}. To our knowledge, there are no known FL security frameworks that would attempt to adapt the security measures holistically which results in insufficient security measures for real world applications of FL with security in mind. Therefore, as we discuss our solution, we will further discuss the need for more holistic approaches to security within FL for analysis of effectiveness of known privacy and integrity measures, computational complexity of such security systems in place, and to further identify potential security threats in complete FL environments. \par

Having that said, in this paper we aim to provide a solution to one of the threats posed by Adaptive Adversaries, specifically targeting the issue of RMCs. As such, the contributions of this paper are as follows:
\begin{enumerate}
    \item Briefly introduce the fundamental elements of FL such that the reader has a concrete understanding of related environments for further security discussions.
    \item Discuss in detail security measures provided by methods using secure aggregation as one of the leading defensive mechanisms for model-robustness.
    \item Provide the reader with relevant understanding of IBI schemes, as well as delve into more detail on the state-of-the-art IBI scheme used in this paper, TNC-IBI.
    \item Present experimental results on IBI usage in FL environments along with the potential benefits it brings, as well as further discuss the surrounding implications for FL security using Identity-Based Cryptography from a holistic perspective of an entire system.
    \item Provide research directions on the security against adaptive adversaries based on the previous discussions, elaborating on concrete pointers and encouraging the community to further explore both issues and defences in the field. 
\end{enumerate}
The rest of this paper is organised as follows. Section 2 will explore related works in the field of FL security, with emphasis on adaptive adversaries and the surrounding topics. Section 3 will provide the reader with preliminary knowledge for Federated Learning, Secure Aggregation, and Identity-Based Identification  covering the necessary background for further discussions. Section 4 will then introduce our solution along with the experimental setup used to produce results, showcasing how have we implemented TNC-IBI over FL system. Section 5 will present the experimental results which will be followed by their detailed discussion in Section 6. We conclude this paper in Section 7 with a concise summary of our findings and contribution.

\section{Related Works}
Adaptive adversaries in context of FL mostly refer to the adaptive modification of attack hyperparameters such that the adversary is capable to conduct their attack stealthily with maximised severity. As there are limited works on the subject, the concrete definition of adaptive adversaries is as of yet missing from the literature. The initial approach proposed in \cite{bagdasaryan_how_2020} focuses on introducing a backdoor, effectively poisoning the global model using  “constrain-and-scale” technique. In essence, the adaptiveness of the attack comes from optimisation for \(\alpha\) hyperparameter, balancing the classification loss and anomaly loss of the poisoned model as presented on the equation below.
\begin{equation}
    L_{model} = \alpha L_{class} + (1-\alpha)L_{anomaly}
\end{equation}
In the authors’ case, the optimisation for \(\alpha\) is done experimentally over multiple malicious clients with general intuition being that its smaller values indicate better detection avoidance and larger values result in increased effectiveness of the attack. \par

Inspired by this approach, a recent improvement to the attack has been presented in \cite{kraus_automatic_2024} introducing AutoAdapt method, which adopts constrain-based optimisation  starting with small values of \(\alpha\) to then gradually maximise the effectiveness of the attack. Doing so allows this method to support multiple constraints as well as work against multi-metric defences, as the value of attack hyperparameter is treated as an optimisation problem. This also partially addresses the concern pointed out by the authors, stating that recent works lack realism in defining the adversaries settings and constraints when evaluating proposed defence methods \cite{kraus_automatic_2024}. \par

Further, in \cite{szelag_adaptive_2025} authors take a different approach to adaptive adversaries and how they are defined. Instead of focusing on specific attack hyperparameters, they assume that an attacker has a capability to change the attack approach entirely if reconnecting clients are not monitored properly. Showing that RMC’s are capable of adapting to the defence methods by changing their own attack even after they have been previously filtered out and disconnected. This is especially relevant given the fact that in current literature there are number of works that are adapted to work against specific defence configurations \cite{zhang_fltracer_2024, hu_defense-guided_2024, gao_secure_2023}, making it a more realistic approach towards simulating adversaries in FL environments. \par

Adaptive adversaries are often omitted when discussing trustworthy AI, their lack of presence in recent review articles such as \cite{tariq_trustworthy_2024} make it apparent that this issue has not gathered as much attention as of yet. That is despite the presence of other works in the direction of adaptive defences \cite{chen_channel_2024}, although after initially surveying the field, adaptiveness in FL is most often addressed in the context of training and communication optimisation \cite{morell_optimising_2022, chai_communication_2023}, with only a handful of works concentrating on exploring this threat and potential defence approaches \cite{zhang_a3fl_2023, kraus_mesas_2023}. \par

Generally, works in related domains which concentrate on attacks and defences for FL systems commonly assume that most of the clients are benign \cite{szelag_adaptive_2025}, operating on a constraint that the maximum number of malicious clients is no higher than \(\lceil \frac{N}{2} \rceil - 1\). The presence of this assumption is especially noticeable in literature covering secure aggregation, including state-of-the-art methods such as FLTrust \cite{cao_fltrust_2022} and FLARE \cite{wang_flare_2022}, as one of the predominant approaches for this defence mechanism uses statistics-based methods that require this constraint to be present. Similarly, attacks against those systems adopt this assumption as a baseline for worst case scenario with the abovementioned maximum quantity of malicious clients \cite{dong_privacy-preserving_2024}. This may be deemed as a realistic setting assuming we are referring to large-scale FL systems, although we note that this might not always be the case and realistic FL environments may may not satisfy that requirement \cite{wang_robust_2022}. 

\section{Preliminaries}
Following this section, our work will propose a defence method that addresses the concern pointed out by \cite{szelag_adaptive_2025} in relation to the threat RMC’s pose against FL systems. As such, our method will ensure that the malicious clients that have been filtered out will not be able to reconnect back in again. For readers convenience, we briefly go through the basic aspects of FL, Secure Aggregation as  the main detection method against model poisoning attacks in FL, and Identity-Based Identification for understanding of our core approach to counter this threat.

\subsection{Federated Learning}
FL is a data decentralized approach to machine learning where the aim is to allow separate clients produce local model using their private datasets. Its core idea revolves around privacy-preservation of separate datasets as the final model produced through FL does not directly learn from the data itself but rather is an amalgamation of local models transferred over to the aggregator or other clients. The FL architecture may vary depending on the needs of the system and heterogeneity of data that the clients own \cite{ye_heterogeneous_2023, martinez_beltran_decentralized_2023}. The following list concisely describes the different types of FL architectures depending on those factors.\par

\leftskip 20pt
\setlength{\parindent}{0em}
\textbf{Horizontal FL.} Clients possess data that contains similar feature space but significantly differs in sample space. It is the most common approach to FL due to the implementation simplicity when compared to other approaches.\par
\textbf{Vertical FL.} Datasets of individual clients differ in the feature space as the sample space is kept similar among clients. Especially applicable in scenarios where the aim is to produce a model over multiple datasets containing different information about the same users.\par
\textbf{Federated Transfer Learning.} When clients datasets possess entirely different feature and sample space. Models adopt transfer learning to leverage insights from the data and address challenges associated with data distribution.  \par
\textbf{Centralised FL.} Characterised with a presence of a central entity, often referred to as an aggregator, that coordinates the FL processes in the system. Clients often are only aware of the aggregator and can function without knowledge of other clients, such architectures are often referred to as server-client. It is the most common type of architecture found in the literature given its simplicity and potential applications. \par
\textbf{Decentralised FL.} Addresses environments where a central entity is not present, which can be caused by limited resources or security concerns. Decentralised FL, also known as peer-to-peer, offer flexibility that come with decentralized architectures and allows for such systems to exist without the requirement of a centralised authority. \par
\textbf{Clustered FL.} Layered approach to FL, where a combination of both approaches, centralised and decentralised, are adopted within the architecture such that the clients are kept within clusters. Besides training individual local models, the system also produces cluster models of each cluster prior to aggregation of the global model. Allows for further flexibility and adaptation of cluster aggregation strategies that address data heterogeneity issues at the cost of system complexity. \par

\vspace{10pt}
\leftskip 0pt
\setlength{\parindent}{2em}

In most cases of FL, the general process of global model training is as follows: (1) global model is first initialised, where all clients are aware of the model weights and associated parameters, (2) clients proceed to train their individual local models based on the only accessible private dataset which is unique to each client, (3) the produced local models are then shared with other clients, the aggregator, or the cluster (depending on the architecture) for aggregation, (4) the global model is aggregated using a chosen aggregation rule that combines all the received local or cluster models, (5) the generated global model is evaluated and sent back to the clients, this process repeats fixed number of times or until a satisfactory accuracy is reached. Every such iteration of the process is commonly referred to as a federated learning round.

\subsection{Secure Aggregation}
The primary goal of Secure Aggregation (SA) is to ensure that during the process of aggregation, the local models which are deemed as unreliable will not be involved in the aggregation \cite{qi_model_2024}. The local models deviations stem from either intentional tampering (model poisoning) or anomalous results, both of which are aimed to be filtered out by SA algorithms. As such, the secure aggregation protocols are most often statistics-based which serve as a baseline for detection of untrustworthy clients. An inherent requirement for effectiveness of SA algorithms is to possess knowledge for general direction of legitimate vectors such that it is able to incorporate detection mechanisms and identify malicious models. Hence the prevalent assumption from Section 2, where majority of clients are assumed to be benign as this allows to utilise legitimate clients for identification of the genuine direction for the global model. \par

Our work will utilise two secure aggregation algorithms commonly used to compare new solutions to: Krum and Trimmed Mean. Despite their simplicity when compared to the other solutions present in the literature, these will effectively present the proof of concept behind our solution.

\subsubsection{Trimmed Mean}
Indiscriminately cuts off statistical deviations at an arbitrary threshold \cite{yin_byzantine-robust_2018}. One chooses a fraction of clients \(\beta\) to be excluded from the aggregation that lies within \([0,\frac{1}{2})\) which will then be excluded on both sides of the  client distribution. In short, \(\beta\) fraction of clients with highest and lowest values will be removed from the aggregation, leaving only the remaining fraction of \(1-2\beta\) clients that remains closest to the mean. As such, given total number of clients \(m\), a set of trimmed clients \(U_x\), we may define each coordinate \(k\) of a vector \(g\) with the following expression \cite{yin_byzantine-robust_2018}:
\begin{equation}
    g_k=\frac{1}{(1-2\beta)m}\sum_{x\in U_x}x
\end{equation}

\subsubsection{Krum}
First introduced for stochastic gradient descent resilience against Byzantine adversaries \cite{blanchard_machine_2017}. The aggregation rule aims to select the most reliable update vector out of \(n\) total clients with \(f\) byzantine clients present using \(n-f-2\) vector updates \(j\) that are neighbouring to each vector \(i\) by calculating their scores \(s(i)\) based on a sum of squared Euclidean distances. The vector with the lowest score value will indicate most reliable update, this may be simply expressed with the following expression:
\begin{equation}
    Krum(V_1,...,V_n) = argmin_i\sum_{i\rightarrow j}||V_i-V_j||^2
\end{equation}

\subsection{Identity-Based Identification}
Identity-Based Identification (IBI) schemes are composed of four probabilistic polynomial-time algorithms such that \(IBI=(SETUP,EXTRACT,PROVE,VERIFY)\), all of which are defined in a following way:

\leftskip 20pt
\setlength{\parindent}{0em}
\textbf{\(SETUP:\)} Using a security parameter \(1^k\) it outputs two master keys, public and secret \((mpk,msk)\).\par
\textbf{\(EXTRACT:\)} Takes master keys \((mpk,msk)\) and user identity \(ID\) to produce a user secret key \(usk\), unique to the provided identity.\par
\textbf{\(PROVE:\)} Takes user secret key \(usk\), user identity \(ID\) and challenge provided by the verifier \(CHA\) and outputs the solution to the challenge. \par
\textbf{\(VERIFY:\)} After receiving the challenge response \(RSP\) from the prover, it takes the commitment instance \(CMT\) and provided identity \(ID\) to decide whether to accept or reject provers response. \par

\vspace{10pt}
\leftskip 0pt
\setlength{\parindent}{2em}

Using the above algorithms, the IBI protocols proceed as follows: (1) Verifier generates a pair of master keys \((mpk,msk)\) using a security parameter \(1^k\), (2) Prover computes their commitment \(CMT\) based on their string-identity \(ID\) mixed with a random salt and sends it to the Verifier, (3) Verifier extracts the user secret key \(usk\) based on the master keys and provided user identity, it also samples a random challenge \(CHA\) (often from predefined finite space of unique challenges), transferring both back to the Prover, (4) Prover computes a solution to the challenge using their user secret key and identity, sending the response \(RSP\) to the Verifier, (5) finally, Verifier checks whether the Provers solution is valid using received response, commitment, and identity, at which point the decision to accept or reject authentication is made. An example of an IBI protocol can be found on Figure 1. \par

\begin{figure*}[ht]
    \centering
    \includegraphics[width=\textwidth]{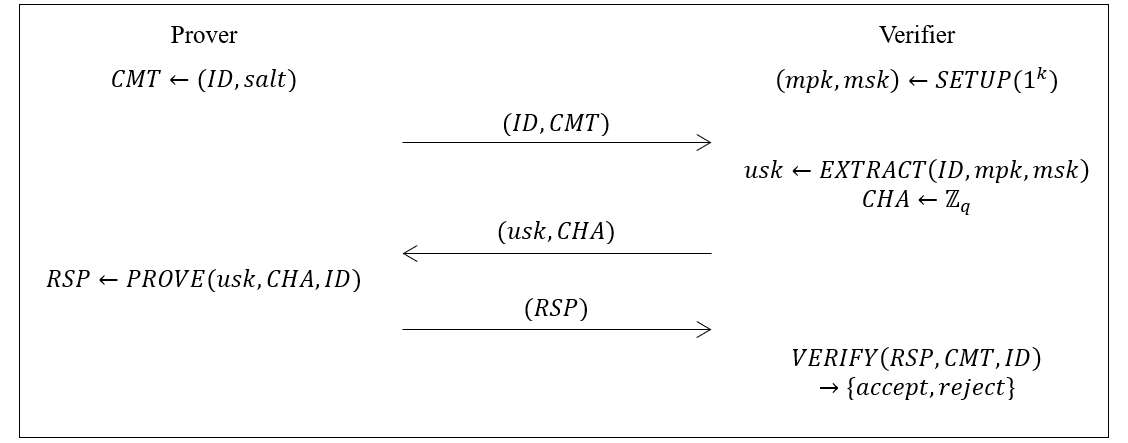}
    \caption{Example IBI protocol.}
    \label{Algorithm}
\end{figure*}

\section{Methodology}
\subsection{Experimental Setup}
In our experiments, we carry out a simulation over a centralized Horizontal FL architecture with 20 clients and one aggregator, implementing a FL environment similar to that from \cite{szelag_adaptive_2025} with the following configuration.  \par

A simple feedforward neural network has been chosen for our global model with \(8*16*8\) fully connected layers using a standard Sigmoid activation function, between which we applied a dropout equal to \(p=0.5\). An additional dropout of \(p=0.2\) has been applied between the last hidden layer and an output layer. Data loaders have been configured with a batch size of 32 and the learning rate was set to \(\alpha = 0.01\). Note that the reason for why our experimental setup strongly resembles that of \cite{szelag_adaptive_2025} is due to the addressed issue at hand; we replicate the same environment that the authors have used to show evidence for the threat of RMC’s and apply our solution over it to prove its effectiveness. \par

Using a similar approach, the simulation of adversarial behaviour proceeds as follows. When adversaries first connect to the FL system, their malicious clients use Gaussian Noise attack \cite{li_experimental_2024} with \(\mu=2\) and \(\sigma=2\) in an attempt to inflict as much damage as possible to the global model. However, if malicious clients adopting this strategy will be detected by secure aggregation algorithms (Krum and Trimmed Mean in two separate sets of experiments) and forcibly disconnected, then malicious clients will proceed to try and reconnect. During adversaries follow-up attempt at poisoning the global model, we implement A Little is Enough (ALIE) \cite{baruch_little_2019} attack with \(z_{max}\) value of 0.9, simulating adversaries attempting to adapt their attack by changing the strategy after being forcibly disconnected, assuming that this knowledge is available to an adversary. Notice how adversarial strategy more focused on stealth of the attack, in simulations presented by \cite{szelag_adaptive_2025} this change from Gaussian Noise to ALIE has resulted in significant drop in accuracy of the global model. \par

The dataset chosen for our simulation were survey responses from Behavrioural Risk Factor Surveillance System (BRFSS 2015) that our model would perform binary classification on for diabetes prediction. The dataset we use from \cite{noauthor_diabetes_2023} comes with cleaned 70692 responses to the survey, serving as a satisfiable sample for our simulations. For the proof of concept, the data is distributed equally among all clients, and we allocate the same amount of computational resources to each client during the simulation. \par

\subsection{Our Solution}
The core of our solution lies with identifying the exact location on which IBI will be placed on. For horizontal FL systems with centralised architectures, it is intuitive that the authentication protocols will be placed on server-client connections. It plays to our advantage that in this scenario the aggregator is a single point of contact that clients are dealing with, and with that in mind, the importance on authenticating each client is therefore put on an aggregator. As such, it is important that the chosen IBI scheme can be implemented without a need for a proxy to authenticate but allows for honest-verifier assumption where the role of a verifier is taken by the aggregator. \par

For this reason, we chose to implement TNC-IBI scheme \cite{chia_pairing-free_2021} which fulfils those requirements. In essence, TNC-IBI takes the TNC signature \cite{ng_variant_2017} and applies Kurosawa-Heng transform such that the signed message m can be used as a verifiable identity \(ID\) required for IBI. In addition, we implement TNC-IBI over NIST256p elliptic curve to account for the resource-constraint setting FL systems are commonly assumed to be placed in (e.g. IoT). \par

The described application of TNC-IBI over FL may be found on Figure 2, and the corresponding command-line outputs after executing our solution may be viewed on Figure 3 and Figure 4. Notice that the main difference in our application over elliptic curves is that when generating \(mpk\), generator \(G,y_1\) and \(y_2\) are distinct points on NIST256p. The rest is reproduced in accordance with the authors scheme. \par

\begin{figure*}[ht]
    \centering
    \includegraphics[width=\textwidth]{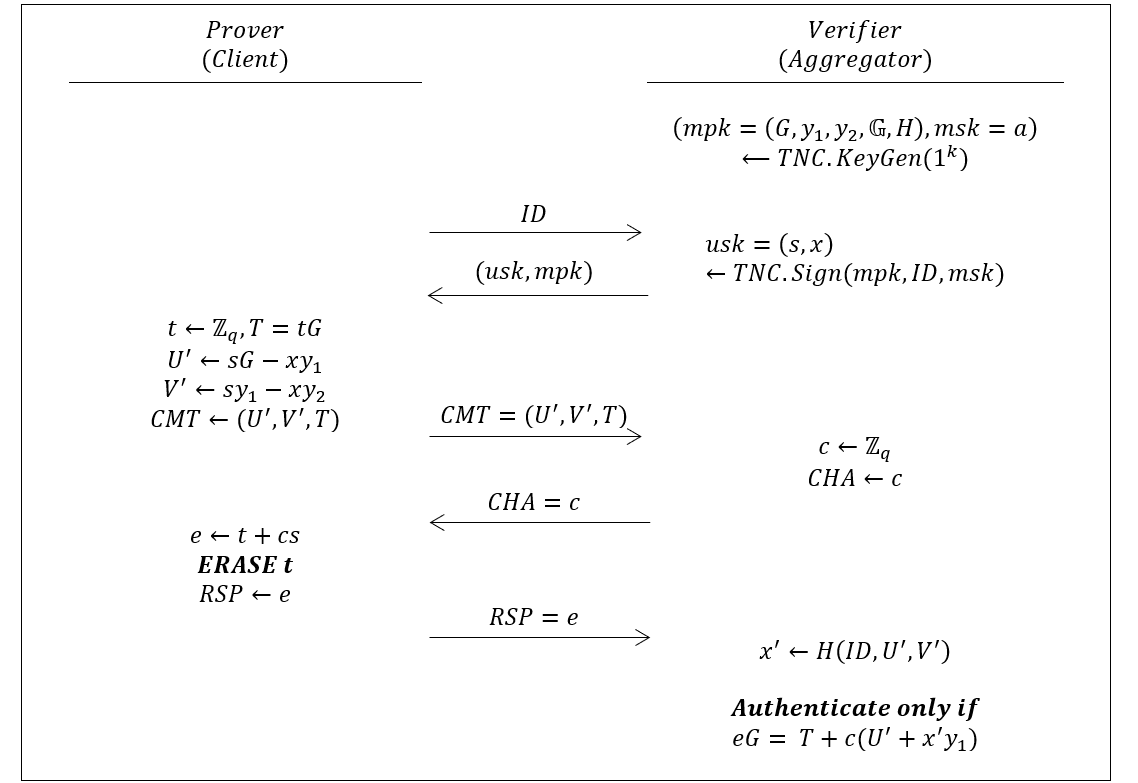}
    \caption{Implementation of our solution reproducing TNC-IBI protocol \cite{chia_pairing-free_2021} over elliptic curves.}
    \label{fig:wide}
\end{figure*}

\begin{figure*}[ht]
    \centering
    \includegraphics[width=\textwidth]{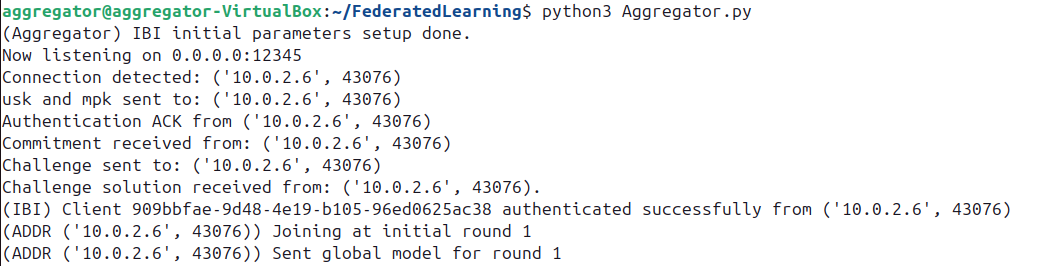}
    \caption{Command-line output for the FL aggregator showcasing parameter generation and authentication control.}
    \label{fig:wide}
\end{figure*}

\begin{figure*}[ht]
    \centering
    \includegraphics[width=\textwidth]{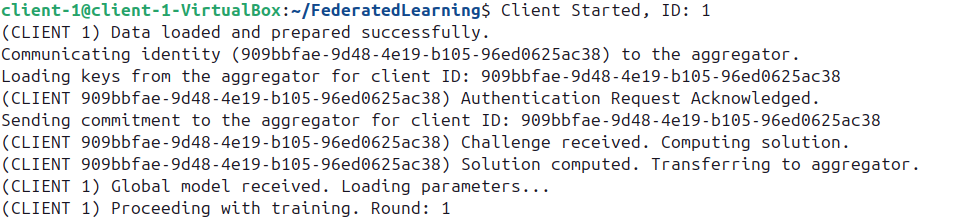}
    \caption{Command-line output for one of the clients showcasing client-side process for authentication in accordance with our proposed solution.}
    \label{fig:wide}
\end{figure*}

\section{Results}
Our results encompass two sets of simulations as it can be seen on Figure 3. In our first round of experiments, we attempted to run simulated RMC’s in an environment with only secure aggregation algorithms (Krum and Trimmed Mean) over 50 iterations. Instead of fixing a point of reconnection like in \cite{szelag_adaptive_2025} we allow for reconnections on separate malicious clients to occur immediately after forcible disconnections. This imitates a more dynamic and active approach taken by the adversary, making the task of associating clients with adversarial activity harder as malicious clients are not initiated in a synchronous manner when reconnecting to the FL environment. \par

Both Krum and Trimmed Mean show to have difficulties in producing accurate results when RMC’s are present, the active behaviour of malicious clients show to significantly degrade overall models accuracy for both of the aggregation rules. Krum reaching an average accuracy of 0.678 with final accuracy of 0.505 and Trimmed Mean an average accuracy of 0.679 with final accuracy of 0.581. In each case both of the final models are deemed as unsatisfactory. Interestingly, Krum managed to allow for partial accuracy persistence during learning, however each time clients were reconnecting with a stealthier attack strategy every following recovery Krum managed to help sustain ended up with a lower accuracy than previously. Trimmed Mean did not achieve any persistency with its results with presence of RMC’s, essentially meaning that every time a RMC’s connected back to the environment the recovered accuracy that Trimmed Mean managed to sustain has again degraded shortly after, rendering the global model unusable. \par

Once our solution has been applied over the FL environment, the IBI authentication managed to retain the identities of malicious clients that have been previously forcibly disconnected. As such, when RMC’s were actively attempting to join back, IBI prevented their authentication preventing further poisoning. This simple, yet straightforward implementation shows that the problem of RMC’s can be effectively prevented using IBI. \par

Combining Krum with TNC-IBI allowed for the global model to achieve final accuracy of 0.74 with consistent training, RMC’s were unable to cause any damage to the global model throughout the entire process as all detected malicious clients were not allowed back into the environment. Similarly, when applying TNC-IBI with Trimmed Mean, the final model was left unpoisoned reaching a final accuracy of 0.746. Although initially, the model did experience some fluctuations which was due to the fact that some malicious clients were not detected from the very beginning as Trimmed Mean aggregation is indiscriminately excludes both ends of the distribution. In which case, if majority of malicious clients were located at one end of the distribution it allowed for a fraction of them to be included in the aggregation. However, as we can see from our results on Figure 3, once the first malicious clients were detected, the rest of RMC’s followed and were permanently excluded from the environment, ensuring that the model reaches consistent results with reliable accuracy. \par

\begin{figure*}[ht]
    \centering
    \includegraphics[width=\textwidth]{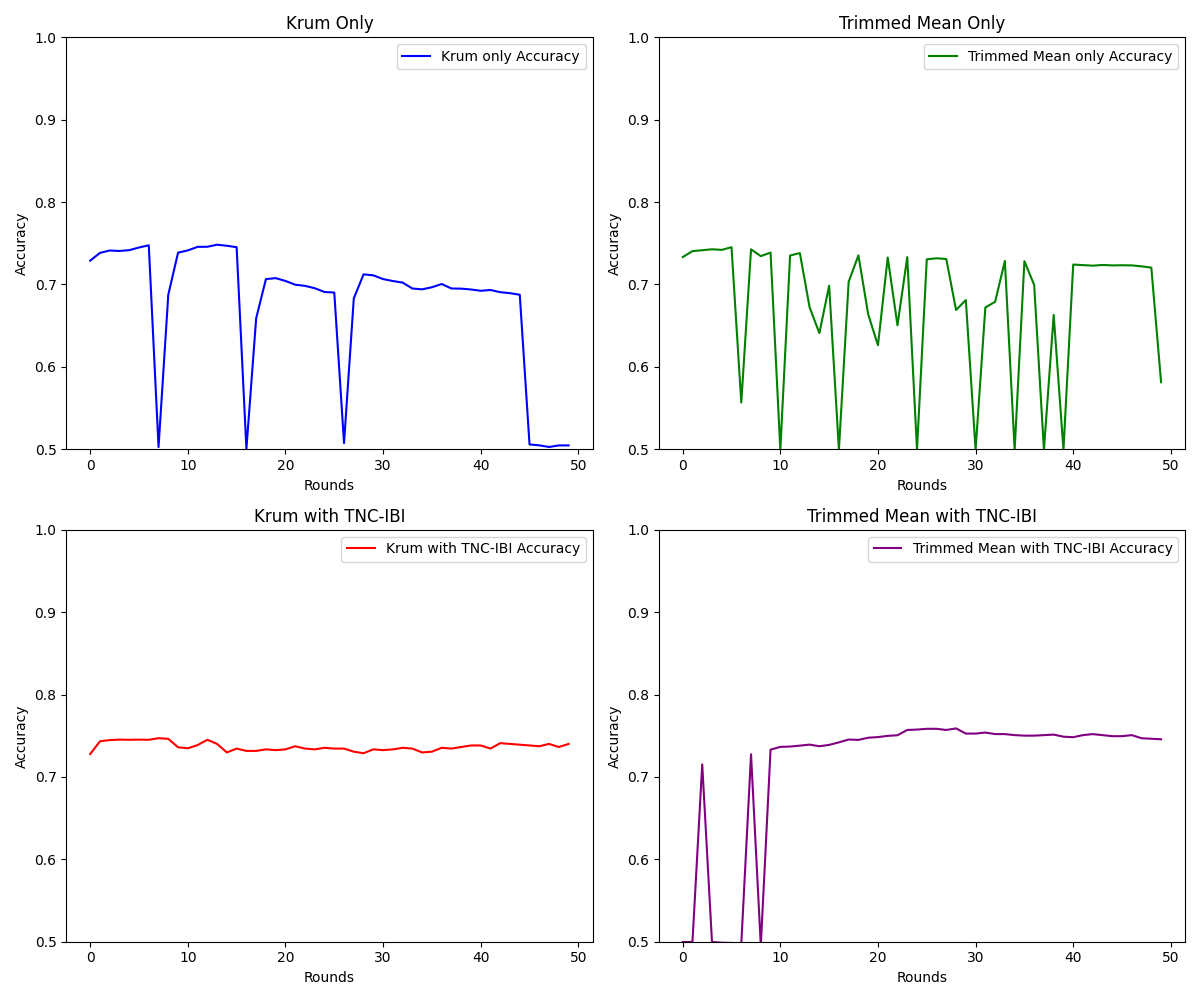}
    \caption{Results for Gaussian Noise and ALIE attacks against two sets of simulations, one including only secure aggregation algorithms, the other with TNC-IBI implement over the FL environment. Showcasing the severity of RMC’s when no defences are present and effectiveness of TNC-IBI as a solution.}
    \label{fig:wide}
\end{figure*}

\section{Research Directions and Discussion}
Now that we have presented the effectiveness of integrating IBI over an FL environment, we now proceed to discuss further implications of our study. \par

\textbf{IBI: Effectiveness.} Our work has shown that usage of IBI in FL is a reasonable option which has not been yet explored in the literature. Acknowledging the fact that the IBI schemes eliminate the need for traditional certificate authorities makes them suitable for FL architectures, so much so that they have potential to regularly aid secure aggregation algorithms in keeping FL systems secure. Both measures compliment one another in their aims which allows their combination to reach a more complete notion for FL security. \par

Nevertheless, with the domain of FL security still in its infancy, much more is yet to be explored. Past works in this domain do not commonly mention use of Identity-Based Cryptography, as such, we are not yet at a point where we can confidently state that IBI is an optimal solution to the issues addressed in this paper. Following our work, we encourage other researchers to further explore this solution and provide further insights as we note that the scope of this paper is limiting the reach of our findings.  \par

\textbf{Reputation Mechanisms.} One of more common methods for determining untrustworthy clients with secure aggregation over time are reputation mechanisms, they allow to determine which clients should be no longer in the environment while avoiding false-negatives by disconnecting clients that are benign. Our work showcases how those or similar mechanisms can support one another in enhancing FL model-robustness. However, that comes with a natural reliance of IBI in our solution on these types of mechanisms as exclusion from further connections may only happen after detection. As such, it is important to further explore how reputation-based secure aggregation algorithms fair against adaptive adversaries when utilising solutions using IBI or future alternatives against adaptive adversaries. We note that even if our theoretical baseline for combining those methods might be sufficient in the future, resource-constraint environments such as IoT may render them incompatible. Hence, given the significance of artificial intelligence and machine learning in the modern world, we call for a proactive approach as we further deepen our understanding in this domain. \par

\textbf{Other FL Architectures.} This paper has strictly concentrated on centralised FL, as such, we highlight that there is a need to address the threat of adaptive adversaries in decentralised and clustered FL architectures. Although it is evident that centralised FL is most commonly used in research, the same cannot be said about real world applications where scenarios that find decentralised FL preferable are much more common than the recent literature presents, e.g. decentralised critical infrastructure \cite{siniosoglou_federated_2024}. Thus, it is equally important to gather attention around other FL architectures and address the same research gaps that we do for centralised FL. \par

\textbf{IoT and Quantum-Safe Efficiency.} Although recently NIST has announced novel standards for post-quantum cryptography such as CRYSTALS-Kyber \cite{bos_crystals_2018} and Dilithium \cite{ducas_crystals-dilithium_2018}, quantum-safe protocols are regularly revisited with intentions of further improvements to security and computational overhead. The latter is often addressed when discussing battery runtime on devices that run these protocols. Recent studies \cite{schoffel_secure_2022} show positive prospects in regard to IoT devices using post-quantum algorithms, showing little difference between some of them and current solutions rooted in Public Key Infrastructure. Nevertheless, the aspect of battery lifetime calls for more improvements when discussing IoT in FL systems, as individual IoT devices that would be a part of FL environments would also be computing local models. In the domain of IoT, computationally demanding tasks such as machine learning stand in the way of novel security implementations and vice versa, due to the fact that IoT environments are very often resource-constraint. Accounting for that interdisciplinary limitation is crucial for progression in this domain to reach implementable, secure FL systems for IoT. \par

\textbf{More on Adaptive Adversaries.} Lastly, our work uses addresses the issue of RMC’s which is a type of adaptive adversary introduced in \cite{szelag_adaptive_2025}. However, it is important to note that this is not the only form of adaptive adversary that can pose threat to FL systems as earlier discussed. Adaptive adversaries in the context of FL are not yet well defined within the scientific literature with only notion being that described in \cite{kraus_mesas_2023} for strong adaptive adversaries. As such, we are not yet fully certain about the extent of the threat adaptive adversaries pose to FL systems. This research gap in particular must be addressed if we want to gain a more meaningful research direction in this domain.

\section{Conclusions}
This paper presents a novel approach to mitigating the threat of Reconnecting Malicious Clients in Federated Learning by employing Identity-Based Identification (IBI) schemes. Through experimental validation, we demonstrate that the integration of IBI with secure aggregation algorithms effectively prevents Reconnecting Malicious Clients (RMCs) from rejoining the system after being detected and forcibly disconnected. Our results indicate that TNC-IBI significantly enhances the security of FL systems by ensuring reliable client authentication without reliance on traditional certificate authorities. Furthermore, we discuss the broader implications of this approach, emphasizing the need for more holistic security frameworks in FL that address adaptive adversaries, reputation-based detection mechanisms, and computational constraints in resource-limited environments. While our work provides a proof-of-concept for IBI in centralized FL, future research should explore its applicability to decentralized architectures and its resilience against emerging adversarial strategies. By addressing these gaps, the research community can move toward more robust, secure, and scalable privacy-preserving machine learning systems.

\bibliographystyle{IEEEtran}
\bibliography{References}

\begin{thebibliography}{10}
\providecommand{\url}[1]{#1}
\csname url@samestyle\endcsname
\providecommand{\newblock}{\relax}
\providecommand{\bibinfo}[2]{#2}
\providecommand{\BIBentrySTDinterwordspacing}{\spaceskip=0pt\relax}
\providecommand{\BIBentryALTinterwordstretchfactor}{4}
\providecommand{\BIBentryALTinterwordspacing}{\spaceskip=\fontdimen2\font plus
\BIBentryALTinterwordstretchfactor\fontdimen3\font minus \fontdimen4\font\relax}
\providecommand{\BIBforeignlanguage}[2]{{%
\expandafter\ifx\csname l@#1\endcsname\relax
\typeout{** WARNING: IEEEtran.bst: No hyphenation pattern has been}%
\typeout{** loaded for the language `#1'. Using the pattern for}%
\typeout{** the default language instead.}%
\else
\language=\csname l@#1\endcsname
\fi
#2}}
\providecommand{\BIBdecl}{\relax}
\BIBdecl

\bibitem{mcmahan_communication-efficient_2017}
\BIBentryALTinterwordspacing
B.~McMahan, E.~Moore, D.~Ramage, S.~Hampson, and B.~A.~y. Arcas, ``\BIBforeignlanguage{en}{Communication-{Efficient} {Learning} of {Deep} {Networks} from {Decentralized} {Data}},'' in \emph{\BIBforeignlanguage{en}{Proceedings of the 20th {International} {Conference} on {Artificial} {Intelligence} and {Statistics}}}.\hskip 1em plus 0.5em minus 0.4em\relax PMLR, Apr. 2017, pp. 1273--1282, iSSN: 2640-3498. [Online]. Available: \url{https://proceedings.mlr.press/v54/mcmahan17a.html}
\BIBentrySTDinterwordspacing

\bibitem{annunziata_dynamics_2024}
\BIBentryALTinterwordspacing
D.~Annunziata, M.~Canzaniello, D.~Chiaro, S.~Izzo, M.~Savoia, and F.~Piccialli, ``On the {Dynamics} of {Non}-{IID} {Data} in {Federated} {Learning} and {High}-{Performance} {Computing},'' in \emph{2024 32nd {Euromicro} {International} {Conference} on {Parallel}, {Distributed} and {Network}-{Based} {Processing} ({PDP})}, Mar. 2024, pp. 230--237, iSSN: 2377-5750. [Online]. Available: \url{https://ieeexplore.ieee.org/document/10495548}
\BIBentrySTDinterwordspacing

\bibitem{zhang_federated_2022}
\BIBentryALTinterwordspacing
T.~Zhang, L.~Gao, C.~He, M.~Zhang, B.~Krishnamachari, and A.~S. Avestimehr, ``Federated {Learning} for the {Internet} of {Things}: {Applications}, {Challenges}, and {Opportunities},'' \emph{IEEE Internet of Things Magazine}, vol.~5, no.~1, pp. 24--29, Mar. 2022, conference Name: IEEE Internet of Things Magazine. [Online]. Available: \url{https://ieeexplore.ieee.org/abstract/document/9773116}
\BIBentrySTDinterwordspacing

\bibitem{elbir_federated_2022}
\BIBentryALTinterwordspacing
A.~M. Elbir, B.~Soner, S.~Çöleri, D.~Gündüz, and M.~Bennis, ``Federated {Learning} in {Vehicular} {Networks},'' in \emph{2022 {IEEE} {International} {Mediterranean} {Conference} on {Communications} and {Networking} ({MeditCom})}, Sep. 2022, pp. 72--77. [Online]. Available: \url{https://ieeexplore.ieee.org/abstract/document/9928621}
\BIBentrySTDinterwordspacing

\bibitem{chaddad_federated_2024}
\BIBentryALTinterwordspacing
A.~Chaddad, Y.~Wu, and C.~Desrosiers, ``Federated {Learning} for {Healthcare} {Applications},'' \emph{IEEE Internet of Things Journal}, vol.~11, no.~5, pp. 7339--7358, Mar. 2024, conference Name: IEEE Internet of Things Journal. [Online]. Available: \url{https://ieeexplore.ieee.org/abstract/document/10288131}
\BIBentrySTDinterwordspacing

\bibitem{nguyen_federated_2022}
\BIBentryALTinterwordspacing
D.~C. Nguyen, Q.-V. Pham, P.~N. Pathirana, M.~Ding, A.~Seneviratne, Z.~Lin, O.~Dobre, and W.-J. Hwang, ``Federated {Learning} for {Smart} {Healthcare}: {A} {Survey},'' \emph{ACM Comput. Surv.}, vol.~55, no.~3, pp. 60:1--60:37, Feb. 2022. [Online]. Available: \url{https://dl.acm.org/doi/10.1145/3501296}
\BIBentrySTDinterwordspacing

\bibitem{castiglia_flexible_2024}
\BIBentryALTinterwordspacing
T.~Castiglia, S.~Wang, and S.~Patterson, ``Flexible {Vertical} {Federated} {Learning} {With} {Heterogeneous} {Parties},'' \emph{IEEE Transactions on Neural Networks and Learning Systems}, vol.~35, no.~12, pp. 17\,878--17\,892, Dec. 2024, conference Name: IEEE Transactions on Neural Networks and Learning Systems. [Online]. Available: \url{https://ieeexplore.ieee.org/abstract/document/10258117}
\BIBentrySTDinterwordspacing

\bibitem{zeng_fedlab_2023}
\BIBentryALTinterwordspacing
D.~Zeng, S.~Liang, X.~Hu, H.~Wang, and Z.~Xu, ``{FedLab}: {A} {Flexible} {Federated} {Learning} {Framework},'' \emph{Journal of Machine Learning Research}, vol.~24, no. 100, pp. 1--7, 2023. [Online]. Available: \url{http://jmlr.org/papers/v24/22-0440.html}
\BIBentrySTDinterwordspacing

\bibitem{xie_federatedscope_2022}
\BIBentryALTinterwordspacing
Y.~Xie, Z.~Wang, D.~Gao, D.~Chen, L.~Yao, W.~Kuang, Y.~Li, B.~Ding, and J.~Zhou, ``{FederatedScope}: {A} {Flexible} {Federated} {Learning} {Platform} for {Heterogeneity},'' Nov. 2022, arXiv:2204.05011 [cs]. [Online]. Available: \url{http://arxiv.org/abs/2204.05011}
\BIBentrySTDinterwordspacing

\bibitem{xie_survey_2024}
\BIBentryALTinterwordspacing
X.~Xie, C.~Hu, H.~Ren, and J.~Deng, ``A survey on vulnerability of federated learning: {A} learning algorithm perspective,'' \emph{Neurocomputing}, vol. 573, p. 127225, Mar. 2024. [Online]. Available: \url{https://www.sciencedirect.com/science/article/pii/S0925231223013486}
\BIBentrySTDinterwordspacing

\bibitem{almutairi_federated_2023}
\BIBentryALTinterwordspacing
S.~Almutairi and A.~Barnawi, ``Federated learning vulnerabilities, threats and defenses: {A} systematic review and future directions,'' \emph{Internet of Things}, vol.~24, p. 100947, Dec. 2023. [Online]. Available: \url{https://www.sciencedirect.com/science/article/pii/S2542660523002706}
\BIBentrySTDinterwordspacing

\bibitem{zhou_deep_2021}
\BIBentryALTinterwordspacing
X.~Zhou, M.~Xu, Y.~Wu, and N.~Zheng, ``\BIBforeignlanguage{en}{Deep {Model} {Poisoning} {Attack} on {Federated} {Learning}},'' \emph{\BIBforeignlanguage{en}{Future Internet}}, vol.~13, no.~3, p.~73, Mar. 2021, number: 3 Publisher: Multidisciplinary Digital Publishing Institute. [Online]. Available: \url{https://www.mdpi.com/1999-5903/13/3/73}
\BIBentrySTDinterwordspacing

\bibitem{liang_egia_2023}
\BIBentryALTinterwordspacing
H.~Liang, Y.~Li, C.~Zhang, X.~Liu, and L.~Zhu, ``{EGIA}: {An} {External} {Gradient} {Inversion} {Attack} in {Federated} {Learning},'' \emph{IEEE Transactions on Information Forensics and Security}, vol.~18, pp. 4984--4995, 2023, conference Name: IEEE Transactions on Information Forensics and Security. [Online]. Available: \url{https://ieeexplore.ieee.org/abstract/document/10209197}
\BIBentrySTDinterwordspacing

\bibitem{zhao_loki_2024}
\BIBentryALTinterwordspacing
J.~C. Zhao, A.~Sharma, A.~R. Elkordy, Y.~H. Ezzeldin, S.~Avestimehr, and S.~Bagchi, ``Loki: {Large}-scale {Data} {Reconstruction} {Attack} against {Federated} {Learning} through {Model} {Manipulation},'' in \emph{2024 {IEEE} {Symposium} on {Security} and {Privacy} ({SP})}, May 2024, pp. 1287--1305, iSSN: 2375-1207. [Online]. Available: \url{https://ieeexplore.ieee.org/abstract/document/10646724}
\BIBentrySTDinterwordspacing

\bibitem{liu_privacy-preserving_2022}
\BIBentryALTinterwordspacing
Z.~Liu, J.~Guo, W.~Yang, J.~Fan, K.-Y. Lam, and J.~Zhao, ``Privacy-{Preserving} {Aggregation} in {Federated} {Learning}: {A} {Survey},'' \emph{IEEE Transactions on Big Data}, pp. 1--20, 2022, conference Name: IEEE Transactions on Big Data. [Online]. Available: \url{https://ieeexplore.ieee.org/abstract/document/9830997}
\BIBentrySTDinterwordspacing

\bibitem{park_privacy-preserving_2022}
\BIBentryALTinterwordspacing
J.~Park and H.~Lim, ``\BIBforeignlanguage{en}{Privacy-{Preserving} {Federated} {Learning} {Using} {Homomorphic} {Encryption}},'' \emph{\BIBforeignlanguage{en}{Applied Sciences}}, vol.~12, no.~2, p. 734, Jan. 2022, number: 2 Publisher: Multidisciplinary Digital Publishing Institute. [Online]. Available: \url{https://www.mdpi.com/2076-3417/12/2/734}
\BIBentrySTDinterwordspacing

\bibitem{ouadrhiri_differential_2022}
\BIBentryALTinterwordspacing
A.~E. Ouadrhiri and A.~Abdelhadi, ``Differential {Privacy} for {Deep} and {Federated} {Learning}: {A} {Survey},'' \emph{IEEE Access}, vol.~10, pp. 22\,359--22\,380, 2022, conference Name: IEEE Access. [Online]. Available: \url{https://ieeexplore.ieee.org/abstract/document/9714350}
\BIBentrySTDinterwordspacing

\bibitem{zhang_uncovering_2024}
\BIBentryALTinterwordspacing
Y.~Zhang, R.~Behnia, A.~A. Yavuz, R.~Ebrahimi, and E.~Bertino, ``Uncovering {Attacks} and {Defenses} in {Secure} {Aggregation} for {Federated} {Deep} {Learning},'' Oct. 2024, arXiv:2410.09676 [cs]. [Online]. Available: \url{http://arxiv.org/abs/2410.09676}
\BIBentrySTDinterwordspacing

\bibitem{kraus_automatic_2024}
\BIBentryALTinterwordspacing
T.~Krauß, J.~König, A.~Dmitrienko, and C.~Kanzow, ``\BIBforeignlanguage{en}{Automatic {Adversarial} {Adaption} for {Stealthy} {Poisoning} {Attacks} in {Federated} {Learning}},'' in \emph{\BIBforeignlanguage{en}{Proceedings 2024 {Network} and {Distributed} {System} {Security} {Symposium}}}.\hskip 1em plus 0.5em minus 0.4em\relax San Diego, CA, USA: Internet Society, 2024. [Online]. Available: \url{https://www.ndss-symposium.org/wp-content/uploads/2024-1366-paper.pdf}
\BIBentrySTDinterwordspacing

\bibitem{bagdasaryan_how_2020}
\BIBentryALTinterwordspacing
E.~Bagdasaryan, A.~Veit, Y.~Hua, D.~Estrin, and V.~Shmatikov, ``\BIBforeignlanguage{en}{How {To} {Backdoor} {Federated} {Learning}},'' in \emph{\BIBforeignlanguage{en}{Proceedings of the {Twenty} {Third} {International} {Conference} on {Artificial} {Intelligence} and {Statistics}}}.\hskip 1em plus 0.5em minus 0.4em\relax PMLR, Jun. 2020, pp. 2938--2948, iSSN: 2640-3498. [Online]. Available: \url{https://proceedings.mlr.press/v108/bagdasaryan20a.html}
\BIBentrySTDinterwordspacing

\bibitem{szelag_adaptive_2025}
\BIBentryALTinterwordspacing
J.~K. Szeląg, J.-J. Chin, and S.-C. Yip, ``Adaptive {Adversaries} in {Byzantine}-{Robust} {Federated} {Learning}: {A} survey.'' 2025, publication info: Preprint. [Online]. Available: \url{https://eprint.iacr.org/2025/510}
\BIBentrySTDinterwordspacing

\bibitem{lora_applying_2024}
\BIBentryALTinterwordspacing
C.~P. Lora, V.~Sandeep, and R.~K. Sinha, ``Applying {Identity}-{Based} {Security} {Algorithms} to {Cloud} {Computing} {Networks},'' in \emph{2024 15th {International} {Conference} on {Computing} {Communication} and {Networking} {Technologies} ({ICCCNT})}, Jun. 2024, pp. 1--6, iSSN: 2473-7674. [Online]. Available: \url{https://ieeexplore.ieee.org/abstract/document/10724265}
\BIBentrySTDinterwordspacing

\bibitem{zhang_security_2022}
\BIBentryALTinterwordspacing
J.~Zhang, H.~Zhu, F.~Wang, J.~Zhao, Q.~Xu, and H.~Li, ``\BIBforeignlanguage{en}{Security and {Privacy} {Threats} to {Federated} {Learning}: {Issues}, {Methods}, and {Challenges}},'' \emph{\BIBforeignlanguage{en}{Security and Communication Networks}}, vol. 2022, no.~1, p. 2886795, 2022, \_eprint: https://onlinelibrary.wiley.com/doi/pdf/10.1155/2022/2886795. [Online]. Available: \url{https://onlinelibrary.wiley.com/doi/abs/10.1155/2022/2886795}
\BIBentrySTDinterwordspacing

\bibitem{wen_survey_2023}
\BIBentryALTinterwordspacing
J.~Wen, Z.~Zhang, Y.~Lan, Z.~Cui, J.~Cai, and W.~Zhang, ``\BIBforeignlanguage{en}{A survey on federated learning: challenges and applications},'' \emph{\BIBforeignlanguage{en}{International Journal of Machine Learning and Cybernetics}}, vol.~14, no.~2, pp. 513--535, Feb. 2023. [Online]. Available: \url{https://doi.org/10.1007/s13042-022-01647-y}
\BIBentrySTDinterwordspacing

\bibitem{zhang_fltracer_2024}
\BIBentryALTinterwordspacing
X.~Zhang, Q.~Liu, Z.~Ba, Y.~Hong, T.~Zheng, F.~Lin, L.~Lu, and K.~Ren, ``{FLTracer}: {Accurate} {Poisoning} {Attack} {Provenance} in {Federated} {Learning},'' \emph{IEEE Transactions on Information Forensics and Security}, vol.~19, pp. 9534--9549, 2024, conference Name: IEEE Transactions on Information Forensics and Security. [Online]. Available: \url{https://ieeexplore.ieee.org/abstract/document/10549523}
\BIBentrySTDinterwordspacing

\bibitem{hu_defense-guided_2024}
C.~Hu, Y.~Liu, M.~Zhang, and Z.~Yang, ``\BIBforeignlanguage{en}{Defense-{Guided} {Adaptive} {Attack} on {Byzantine}-{Robust} {Federated} {Learning}},'' in \emph{\BIBforeignlanguage{en}{Frontiers in {Cyber} {Security}}}, B.~Chen, X.~Fu, and M.~Huang, Eds.\hskip 1em plus 0.5em minus 0.4em\relax Singapore: Springer Nature, 2024, pp. 105--119.

\bibitem{gao_secure_2023}
\BIBentryALTinterwordspacing
J.~Gao, B.~Hou, X.~Guo, Z.~Liu, Y.~Zhang, K.~Chen, and J.~Li, ``Secure {Aggregation} is {Insecure}: {Category} {Inference} {Attack} on {Federated} {Learning},'' \emph{IEEE Transactions on Dependable and Secure Computing}, vol.~20, no.~1, pp. 147--160, Jan. 2023, conference Name: IEEE Transactions on Dependable and Secure Computing. [Online]. Available: \url{https://ieeexplore.ieee.org/abstract/document/9618806}
\BIBentrySTDinterwordspacing

\bibitem{tariq_trustworthy_2024}
\BIBentryALTinterwordspacing
A.~Tariq, M.~A. Serhani, F.~M. Sallabi, E.~S. Barka, T.~Qayyum, H.~M. Khater, and K.~A. Shuaib, ``Trustworthy {Federated} {Learning}: {A} {Comprehensive} {Review}, {Architecture}, {Key} {Challenges}, and {Future} {Research} {Prospects},'' \emph{IEEE Open Journal of the Communications Society}, vol.~5, pp. 4920--4998, 2024, conference Name: IEEE Open Journal of the Communications Society. [Online]. Available: \url{https://ieeexplore.ieee.org/document/10623386?arnumber=10623386}
\BIBentrySTDinterwordspacing

\bibitem{chen_channel_2024}
\BIBentryALTinterwordspacing
Z.~Chen, J.~Du, X.~Hou, K.~Yu, J.~Wang, and Z.~Han, ``Channel {Adaptive} and {Sparsity} {Personalized} {Federated} {Learning} for {Privacy} {Protection} in {Smart} {Healthcare} {Systems},'' \emph{IEEE Journal of Biomedical and Health Informatics}, vol.~28, no.~6, pp. 3248--3257, Jun. 2024, conference Name: IEEE Journal of Biomedical and Health Informatics. [Online]. Available: \url{https://ieeexplore.ieee.org/abstract/document/10399784}
\BIBentrySTDinterwordspacing

\bibitem{morell_optimising_2022}
J.~{\'A}. Morell, Z.~A. Dahi, F.~Chicano, G.~Luque, and E.~Alba, ``\BIBforeignlanguage{en}{Optimising {Communication} {Overhead} in {Federated} {Learning} {Using} {NSGA}-{II}},'' in \emph{\BIBforeignlanguage{en}{Applications of {Evolutionary} {Computation}}}, J.~L. Jiménez~Laredo, J.~I. Hidalgo, and K.~O. Babaagba, Eds.\hskip 1em plus 0.5em minus 0.4em\relax Cham: Springer International Publishing, 2022, pp. 317--333.

\bibitem{chai_communication_2023}
\BIBentryALTinterwordspacing
Z.-y. Chai, C.-d. Yang, and Y.-l. Li, ``\BIBforeignlanguage{en}{Communication efficiency optimization in federated learning based on multi-objective evolutionary algorithm},'' \emph{\BIBforeignlanguage{en}{Evolutionary Intelligence}}, vol.~16, no.~3, pp. 1033--1044, Jun. 2023. [Online]. Available: \url{https://doi.org/10.1007/s12065-022-00718-x}
\BIBentrySTDinterwordspacing

\bibitem{zhang_a3fl_2023}
\BIBentryALTinterwordspacing
H.~Zhang, J.~Jia, J.~Chen, L.~Lin, and D.~Wu, ``\BIBforeignlanguage{en}{{A3FL}: {Adversarially} {Adaptive} {Backdoor} {Attacks} to {Federated} {Learning}},'' \emph{\BIBforeignlanguage{en}{Advances in Neural Information Processing Systems}}, vol.~36, pp. 61\,213--61\,233, Dec. 2023. [Online]. Available: \url{https://proceedings.neurips.cc/paper_files/paper/2023/hash/c07d71ff0bc042e4b9acd626a79597fa-Abstract-Conference.html}
\BIBentrySTDinterwordspacing

\bibitem{kraus_mesas_2023}
\BIBentryALTinterwordspacing
T.~Krauß and A.~Dmitrienko, ``\BIBforeignlanguage{en}{{MESAS}: {Poisoning} {Defense} for {Federated} {Learning} {Resilient} against {Adaptive} {Attackers}},'' in \emph{\BIBforeignlanguage{en}{Proceedings of the 2023 {ACM} {SIGSAC} {Conference} on {Computer} and {Communications} {Security}}}.\hskip 1em plus 0.5em minus 0.4em\relax Copenhagen Denmark: ACM, Nov. 2023, pp. 1526--1540. [Online]. Available: \url{https://dl.acm.org/doi/10.1145/3576915.3623212}
\BIBentrySTDinterwordspacing

\bibitem{cao_fltrust_2022}
\BIBentryALTinterwordspacing
X.~Cao, M.~Fang, J.~Liu, and N.~Z. Gong, ``{FLTrust}: {Byzantine}-robust {Federated} {Learning} via {Trust} {Bootstrapping},'' Apr. 2022, arXiv:2012.13995 [cs]. [Online]. Available: \url{http://arxiv.org/abs/2012.13995}
\BIBentrySTDinterwordspacing

\bibitem{wang_flare_2022}
\BIBentryALTinterwordspacing
N.~Wang, Y.~Xiao, Y.~Chen, Y.~Hu, W.~Lou, and Y.~T. Hou, ``{FLARE}: {Defending} {Federated} {Learning} against {Model} {Poisoning} {Attacks} via {Latent} {Space} {Representations},'' in \emph{Proceedings of the 2022 {ACM} on {Asia} {Conference} on {Computer} and {Communications} {Security}}, ser. {ASIA} {CCS} '22.\hskip 1em plus 0.5em minus 0.4em\relax New York, NY, USA: Association for Computing Machinery, May 2022, pp. 946--958. [Online]. Available: \url{https://dl.acm.org/doi/10.1145/3488932.3517395}
\BIBentrySTDinterwordspacing

\bibitem{dong_privacy-preserving_2024}
\BIBentryALTinterwordspacing
C.~Dong, J.~Weng, M.~Li, J.-N. Liu, Z.~Liu, Y.~Cheng, and S.~Yu, ``Privacy-{Preserving} and {Byzantine}-{Robust} {Federated} {Learning},'' \emph{IEEE Transactions on Dependable and Secure Computing}, vol.~21, no.~2, pp. 889--904, Mar. 2024, conference Name: IEEE Transactions on Dependable and Secure Computing. [Online]. Available: \url{https://ieeexplore.ieee.org/document/10093038}
\BIBentrySTDinterwordspacing

\bibitem{wang_robust_2022}
\BIBentryALTinterwordspacing
X.~Wang, K.~Nahrstedt, and O.~O. Koyejo, ``\BIBforeignlanguage{en}{Robust {Federated} {Learning} with {Majority} {Adversaries} via {Projection}-based {Re}-weighting},'' Sep. 2022. [Online]. Available: \url{https://openreview.net/forum?id=adgYjVvm9Xy}
\BIBentrySTDinterwordspacing

\bibitem{ye_heterogeneous_2023}
\BIBentryALTinterwordspacing
M.~Ye, X.~Fang, B.~Du, P.~C. Yuen, and D.~Tao, ``Heterogeneous {Federated} {Learning}: {State}-of-the-art and {Research} {Challenges},'' \emph{ACM Comput. Surv.}, vol.~56, no.~3, pp. 79:1--79:44, Oct. 2023. [Online]. Available: \url{https://dl.acm.org/doi/10.1145/3625558}
\BIBentrySTDinterwordspacing

\bibitem{martinez_beltran_decentralized_2023}
\BIBentryALTinterwordspacing
E.~T. Martínez~Beltrán, M.~Q. Pérez, P.~M.~S. Sánchez, S.~L. Bernal, G.~Bovet, M.~G. Pérez, G.~M. Pérez, and A.~H. Celdrán, ``Decentralized {Federated} {Learning}: {Fundamentals}, {State} of the {Art}, {Frameworks}, {Trends}, and {Challenges},'' \emph{IEEE Communications Surveys \& Tutorials}, vol.~25, no.~4, pp. 2983--3013, 2023, conference Name: IEEE Communications Surveys \& Tutorials. [Online]. Available: \url{https://ieeexplore.ieee.org/abstract/document/10251949}
\BIBentrySTDinterwordspacing

\bibitem{qi_model_2024}
\BIBentryALTinterwordspacing
P.~Qi, D.~Chiaro, A.~Guzzo, M.~Ianni, G.~Fortino, and F.~Piccialli, ``Model aggregation techniques in federated learning: {A} comprehensive survey,'' \emph{Future Generation Computer Systems}, vol. 150, pp. 272--293, Jan. 2024. [Online]. Available: \url{https://www.sciencedirect.com/science/article/pii/S0167739X23003333}
\BIBentrySTDinterwordspacing

\bibitem{yin_byzantine-robust_2018}
\BIBentryALTinterwordspacing
D.~Yin, Y.~Chen, R.~Kannan, and P.~Bartlett, ``\BIBforeignlanguage{en}{Byzantine-{Robust} {Distributed} {Learning}: {Towards} {Optimal} {Statistical} {Rates}},'' in \emph{\BIBforeignlanguage{en}{Proceedings of the 35th {International} {Conference} on {Machine} {Learning}}}.\hskip 1em plus 0.5em minus 0.4em\relax PMLR, Jul. 2018, pp. 5650--5659, iSSN: 2640-3498. [Online]. Available: \url{https://proceedings.mlr.press/v80/yin18a.html}
\BIBentrySTDinterwordspacing

\bibitem{blanchard_machine_2017}
\BIBentryALTinterwordspacing
P.~Blanchard, E.~M. El~Mhamdi, R.~Guerraoui, and J.~Stainer, ``Machine {Learning} with {Adversaries}: {Byzantine} {Tolerant} {Gradient} {Descent},'' in \emph{Advances in {Neural} {Information} {Processing} {Systems}}, vol.~30.\hskip 1em plus 0.5em minus 0.4em\relax Curran Associates, Inc., 2017. [Online]. Available: \url{https://proceedings.neurips.cc/paper_files/paper/2017/hash/f4b9ec30ad9f68f89b29639786cb62ef-Abstract.html}
\BIBentrySTDinterwordspacing

\bibitem{li_experimental_2024}
\BIBentryALTinterwordspacing
S.~Li, E.~C.-H. Ngai, and T.~Voigt, ``An {Experimental} {Study} of {Byzantine}-{Robust} {Aggregation} {Schemes} in {Federated} {Learning},'' \emph{IEEE Transactions on Big Data}, vol.~10, no.~6, pp. 975--988, Dec. 2024, conference Name: IEEE Transactions on Big Data. [Online]. Available: \url{https://ieeexplore.ieee.org/abstract/document/10018261}
\BIBentrySTDinterwordspacing

\bibitem{baruch_little_2019}
\BIBentryALTinterwordspacing
G.~Baruch, M.~Baruch, and Y.~Goldberg, ``A {Little} {Is} {Enough}: {Circumventing} {Defenses} {For} {Distributed} {Learning},'' in \emph{Advances in {Neural} {Information} {Processing} {Systems}}, vol.~32.\hskip 1em plus 0.5em minus 0.4em\relax Curran Associates, Inc., 2019. [Online]. Available: \url{https://proceedings.neurips.cc/paper_files/paper/2019/hash/ec1c59141046cd1866bbbcdfb6ae31d4-Abstract.html}
\BIBentrySTDinterwordspacing

\bibitem{noauthor_diabetes_2023}
\BIBentryALTinterwordspacing
``\BIBforeignlanguage{en}{Diabetes, {Hypertension} and {Stroke} {Prediction}},'' 2023. [Online]. Available: \url{https://www.kaggle.com/datasets/prosperchuks/health-dataset}
\BIBentrySTDinterwordspacing

\bibitem{chia_pairing-free_2021}
\BIBentryALTinterwordspacing
J.~Chia, J.-J. Chin, and S.-C. Yip, ``\BIBforeignlanguage{en}{A {Pairing}-{Free} {Identity}-{Based} {Identification} {Scheme} with {Tight} {Security} {Using} {Modified}-{Schnorr} {Signatures}},'' \emph{\BIBforeignlanguage{en}{Symmetry}}, vol.~13, no.~8, p. 1330, Aug. 2021, number: 8 Publisher: Multidisciplinary Digital Publishing Institute. [Online]. Available: \url{https://www.mdpi.com/2073-8994/13/8/1330}
\BIBentrySTDinterwordspacing

\bibitem{ng_variant_2017}
\BIBentryALTinterwordspacing
T.-S. Ng, S.-Y. Tan, and J.-J. Chin, ``A variant of {Schnorr} signature scheme with tight security reduction,'' in \emph{2017 {International} {Conference} on {Information} and {Communication} {Technology} {Convergence} ({ICTC})}, Oct. 2017, pp. 411--415. [Online]. Available: \url{https://ieeexplore.ieee.org/abstract/document/8191014}
\BIBentrySTDinterwordspacing

\bibitem{siniosoglou_federated_2024}
\BIBentryALTinterwordspacing
I.~Siniosoglou, S.~Bibi, K.-F. Kollias, G.~Fragulis, P.~Radoglou-Grammatikis, T.~Lagkas, V.~Argyriou, V.~Vitsas, and P.~Sarigiannidis, ``\BIBforeignlanguage{eng}{Federated {Learning} {Models} in {Decentralized} {Critical} {Infrastructure}},'' in \emph{\BIBforeignlanguage{eng}{Shaping the {Future} of {IoT} with {Edge} {Intelligence}: {How} {Edge} {Computing} {Enables} the {Next} {Generation} of {IoT} {Applications}}}, ser. Taylor \& {Francis} {Collection} of {European} {Union} {Research} and {Innovation} {Funded} {Monographs} and {Chapters}, R.~C. Sofia and J.~Soldatos, Eds.\hskip 1em plus 0.5em minus 0.4em\relax Abingdon (UK): River Publishers, 2024, pp. 95--115. [Online]. Available: \url{http://www.ncbi.nlm.nih.gov/books/NBK602372/}
\BIBentrySTDinterwordspacing

\bibitem{bos_crystals_2018}
\BIBentryALTinterwordspacing
J.~Bos, L.~Ducas, E.~Kiltz, T.~Lepoint, V.~Lyubashevsky, J.~M. Schanck, P.~Schwabe, G.~Seiler, and D.~Stehle, ``{CRYSTALS} - {Kyber}: {A} {CCA}-{Secure} {Module}-{Lattice}-{Based} {KEM},'' in \emph{2018 {IEEE} {European} {Symposium} on {Security} and {Privacy} ({EuroS}\&{P})}, Apr. 2018, pp. 353--367. [Online]. Available: \url{https://ieeexplore.ieee.org/document/8406610}
\BIBentrySTDinterwordspacing

\bibitem{ducas_crystals-dilithium_2018}
\BIBentryALTinterwordspacing
L.~Ducas, E.~Kiltz, T.~Lepoint, V.~Lyubashevsky, P.~Schwabe, G.~Seiler, and D.~Stehlé, ``\BIBforeignlanguage{en}{{CRYSTALS}-{Dilithium}: {A} {Lattice}-{Based} {Digital} {Signature} {Scheme}},'' \emph{\BIBforeignlanguage{en}{IACR Transactions on Cryptographic Hardware and Embedded Systems}}, pp. 238--268, Feb. 2018. [Online]. Available: \url{https://tches.iacr.org/index.php/TCHES/article/view/839}
\BIBentrySTDinterwordspacing

\bibitem{schoffel_secure_2022}
\BIBentryALTinterwordspacing
M.~Schöffel, F.~Lauer, C.~C. Rheinländer, and N.~Wehn, ``\BIBforeignlanguage{en}{Secure {IoT} in the {Era} of {Quantum} {Computers}—{Where} {Are} the {Bottlenecks}?}'' \emph{\BIBforeignlanguage{en}{Sensors}}, vol.~22, no.~7, p. 2484, Jan. 2022, number: 7 Publisher: Multidisciplinary Digital Publishing Institute. [Online]. Available: \url{https://www.mdpi.com/1424-8220/22/7/2484}
\BIBentrySTDinterwordspacing

\end{thebibliography}

\end{document}